\documentclass[%
reprint,
 amsmath,amssymb,
 aps,
prl,
]{revtex4-1}

\usepackage{lipsum}
\usepackage{graphicx}
\usepackage{bm}
\usepackage{color}
\usepackage{times}
\usepackage{amsmath}
\usepackage{float}
\usepackage{hyperref}
\usepackage{ulem} 
\usepackage[dvipsnames]{xcolor}

\newcommand{\beginsupplement}{%
	\setcounter{table}{0}
	\renewcommand{\thetable}{S\arabic{table}}%
	\setcounter{figure}{0}
	\renewcommand{\thefigure}{S\arabic{figure}}%
}

\begin{document}

\preprint{APS/123-QED}

\title{Exploiting Localized Transition Waves to Tune Sound Propagation in Soft Materials}

\author{Audrey A. Watkins$^{\dagger}$, Austin Eichelberg$^\dagger$} 
\author{Osama R. Bilal} \email{osama.bilal@uconn.edu\\ $^\dagger$ Both authors contributed equally to this work.}

\affiliation{Department of Mechanical engineering, University of Connecticut, Storrs, CT 06269, USA}
\date{\today}

\begin{abstract}

Programmable materials hold great potential for many applications such as deployable structures, soft robotics, and wave control, however, the presence of instability and disorder might hinder their utilization. Through a combination of analytical, numerical, and experimental analyses, we harness the interplay between instabilities, geometric frustration,  and mechanical deformations to control the propagation of sound waves within self-assembled soft materials. We consider levitated magnetic disks confined by a magnetic boundary \textcolor{black}{in-plane}. The assemblies can be either ordered or disordered depending on the intrinsic disk symmetry. By applying an external load to the assembly, we observe the nucleation and propagation of different topological defects within the lattices. In the presence of instabilities, the defect propagation gives rise to  \textit{time-independent} localized transition waves. Surprisingly, in the presence of frustration, the applied load briefly introduces \textit{deformation-induced order} to the material. By further deforming the lattices, new patterns emerge across all disk symmetries. We utilize these patterns to tune sound propagation through the material. Our findings could open new possibilities for designing exotic materials with potential applications ranging from sound control to soft robotics.

\end{abstract}

\maketitle


Patterns emerge within an assembly when the system elements are left to interact based on their individual energy. When an external load is applied to these self-assembled patterns, the elements keep reorienting themselves to minimize their energy, resulting in different patterns\cite{taheri2015self}. In the case of incompatible elements, geometric frustration\cite{mellado2012macroscopic,schonke2015infinite,wang2017harnessing,kang2014complex} can arise, particularly, when the global energy minima of the system is incompatible with the minima of the individual elements\cite{diep2013frustrated}. Such geometric frustration can prohibit the self-assembly from achieving long-range order. Furthermore, the presence of instabilities can influence the emerging order (or disorder) within a tunable material\cite{kochmann2017exploiting,li2019domain,goshkoderia2020instability,jin2020guided,haghpanah2016multistable,florijn2014programmable,florijn2016programmable}. \textcolor{black}{A valid path for functional tunability of matter is external stimuli (such as heat or magnetic fields) which can effectively apply an external load to tune material properties \cite{matar2013tunable,bilal2017bistable,bilal2017reprogrammable,palermo2019tuning,wang2018observation,wang2020tunable}. For soft materials, however, this usually translates to large deformation with the potential rise of instabilities, disorder, incompatibility and geometric frustration. While such materials hold great potential for many applications such as deployable structures, soft robotics\cite{chen2018harnessing,rothemund2018soft,zareei2020harnessing} and wave control\cite{bertoldi2008wave,wang2013effects,shan2014harnessing,shim2015harnessing,raney2016stable,bilal2017bistable,bilal2017reprogrammable,li2012switching}, the presence of frustration, instability and disorder within tunable soft materials might hinder their utilization.} Here, we analytically, numerically, and experimentally show that in the presence of instabilities and disorder, pattern evolution due to deformation can be harnessed. By deforming our lattices, we control the nucleation and propagation of \textcolor{black}{topological} defects as time-independent localized transition waves. Furthermore, we exploit the evolving patterns to tune the propagation of sound waves within our material systems.




\begin{figure}[b!]
\begin{center}
\includegraphics{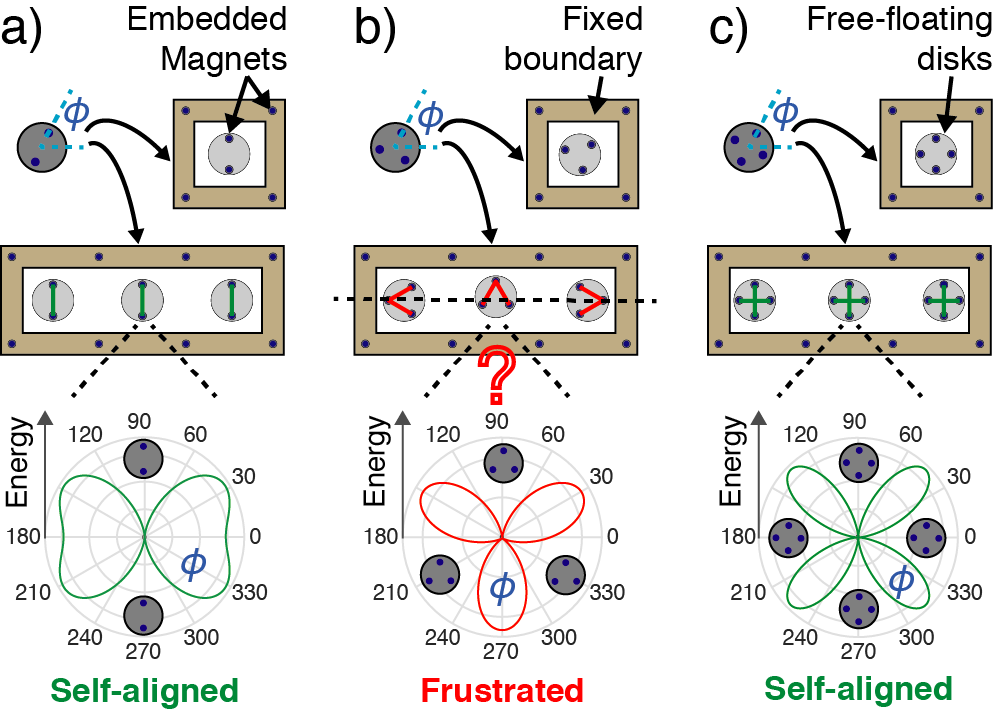}
\caption{\label{fig:concept} \textbf{Concept.} Minimum energy state for 1 disk (top) and 3 disks (middle) confined by a fixed boundary with (a) two, (b) three and (c) four magnets per disk. The potential energy as a function of the disk phase-angle for the central disk is shown as a polar plot (bottom).}
\end{center}
\end{figure}

We consider the mechanics of levitated disks \textcolor{black}{(lying in the horizontal plane)} with repulsive magnetic interactions within a fixed magnetic boundary. The boundary magnets are equally spaced. The embedded disk magnets vary between 1-4, resembling the shape of a dot, line, triangle and a square (Fig.\ref{fig:concept}). The boundary magnets create energy potential-wells dictating the alignment of the free-floating disks. When confined within an energy potential well, a single disk translates and rotates freely to minimize its own energy. For a disk with one magnet, the minimum energy state is at the center of the energy potential well (i.e., the center of the four boundary magnets). In the case of two or four magnets per disk, the disk has four degenerate minimum energy states located at the center of the potential well with a 90$^{\circ}$ phase angle $\phi$ (Fig.\ref{fig:concept}a,c). In the case of three magnets per disk, there exist multiple minimum energy states that are off-center. Each of these off-center positions has three minimum energy states with a 120$^{\circ}$ phase angle (Fig.\ref{fig:concept}b). When multiple disks are confined within the boundary, the self-alignment must take into account the added nearest neighbour interaction. For both two and four magnets per disk, the minimum energy states exist at the center of the potential wells, resulting in perfectly aligned self-oriented lattices (Fig.\ref{fig:concept}a,c middle). In the case of three magnets per disk, the disks have no clear state of self-alignment that minimizes both their individual and total energies, hence the system is \textit{frustrated} (Fig.\ref{fig:concept}c). To experimentally validate the potential energy calculations (Fig.\ref{fig:concept}a-c), we enclose each disk type (i.e., with 1-4 magnets) within a fixed boundary \textcolor{black}{atop an air bearing table}. Both examples of the numerical and the experimental self-aligned disks with 1-4 embedded magnets are included in the Supplementary Material. The disks start in random positions relative to the boundary and to each other. Then, we \textcolor{black}{pressurize the air bearing, activating} a layer of laminar air flow beneath the surface of the disks to allow them to float freely \textcolor{black}{in the horizontal plane} (i.e., levitating the disks similar to the arcade game ``Air Hockey").

\begin{figure}
\begin{center}
\includegraphics[width = \columnwidth]{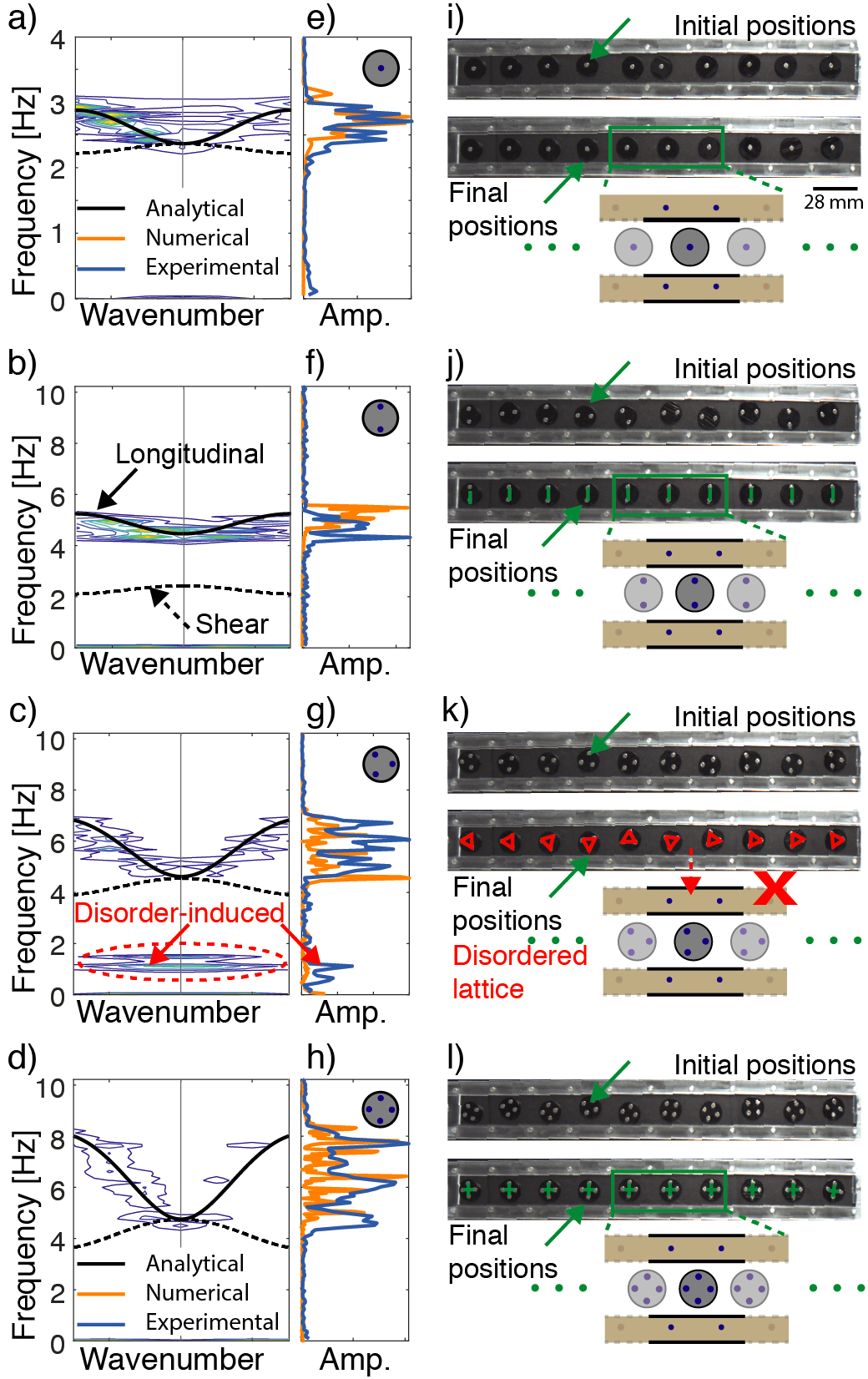}%
\caption{\label{fig:Dispersion} \textbf{Lattice wave propagation}. (a-d) Analytical (black) and experimental (contour) dispersion curves for 1-4 magnets per disk. Numerical (orange) and experimental (blue) transmission at a central disk within the assembly. (i-l) The initial and final positions of ten disks with 1-4 embedded magnets. All the lattices are perfectly aligned except for 3 magnets per disk in panel (k).}
\end{center}
\end{figure}

To study the wave propagation characteristics of the assemblies, we first consider a model of infinite periodic lattices with a single disk and its four boundary magnets as the unit cell with nearest neighbour interaction (i.e., the immediately adjacent disks). \textcolor{black}{Only the free floating disks are allowed to move on both sides of the considered disk.} Each disk has two degrees of freedom in the $x$ and $y$ directions.  The dispersion relation of the system can be calculated using Bloch's theorem\cite{bloch1929quantenmechanik} as: $[-\omega^2\textbf{M}+\textbf{K}(\boldsymbol\kappa)] \boldsymbol{\phi} = 0,$ where $\omega$ is the frequency, $\kappa$ is the wavenumber, $\boldsymbol\phi = [u~v]^{T}$ is the Bloch displacement vector in the $x$ and $y$ directions, $\boldsymbol {M}$ is the mass matrix, and $\boldsymbol {K(\kappa)}$ is the stiffness matrix\cite{watkins2020demultiplexing}, taking into consideration the  static repulsive forces between the disks in the equilibrium configuration\cite{jiao2020dynamics} (See Supp. Mat.). For disks with three embedded magnets, we assume perfect alignment of the disks in order to use Bloch's theorem. The analytically computed dispersion curves show two distinctive bands representing the longitudinal and shear motion in each of the four lattices (Fig.\ref{fig:Dispersion}a-d). In addition, as the number of magnets per disk increases, the frequency of the bands increases, due to the increase in stiffness within the lattice.

To numerically verify the infinite model, we consider a finite arrangement of 10 disks confined to a magnetic boundary (using Verlet method\cite{press2007numerical}). The disks can move freely in-plane. The simulations are initialized with random disk positions. After the model reaches equilibrium, based on the balance of repulsive forces, we excite the right most disk with a chirp signal between 0.2 and 10 Hz.  The fast Fourier transform (FFT) of the longitudinal transmission ranges match closely with the analytical model (Fig.\ref{fig:Dispersion}e-h). Remarkably, in the case of frustrated disks (i.e., with 3 magnets), the simulated transmission through the disordered lattice closely matches the dispersion curve calculated for a perfectly ordered lattice.

To experimentally verify the dispersion curves, we harmonically excite the \textcolor{black}{longitudinal mode of the} floating self-aligned lattices (Fig.\ref{fig:Dispersion}i-l) using a mechanical shaker with a chirp signal between 0.2 - 10 Hz and measure the x-displacement of the disks through digital image correlation (DICe). \textcolor{black}{The longitudinal excitation takes place at the right most disk. The oscillatory motion of the disks is processed using 1D and 2D fast Fourier transform, resulting in a transmission spectrum and an experimental dispersion curve, respectively (See Supp. Mat.).}

The experimentally measured transmission ranges (1D-FFT) and dispersion curves (2D-FFT) are in agreement with our numerical predictions \textcolor{black}{within the longitudinal mode} (Fig.\ref{fig:Dispersion}e-h). Once more, in presence of frustration, the experimentally measured transmission and dispersion through the disordered lattice matches closely with a perfectly ordered lattice. In addition, we note the emergence of an extra transmission band around 1 Hz for both measured transmission and dispersion. This transmission anomaly dissipates as the wave propagates further down the lattice due to the increased disorder. \textcolor{black}{We note that there is no experimental transmission correlating with the shear modes, due to the absence of shear excitation.}


A viable strategy for tuning wave propagation in soft material is applying an external load to tune stiffness\cite{boechler2011tunable,wang2014harnessing,florijn2014programmable,babaee2016harnessing,amendola2018tuning}, however, this usually tunes the material globally. An intriguing feature of our material is the localized pattern transformation enabled by the boundary-induced multi-stability of the potential wells. When a load is applied at one end of the material, the disks initially experience an added global stiffness. Once the applied load passes a threshold, the first disk overcomes its boundary potential and snaps into the next potential well. As the compression at the boundary increases, the migration of disks between potentials creates a propagating defect, which gives rise to a stable transition wave. If the compression pauses, the propagation of the defects throughout the lattice stops (i.e., the transition wave is \textit{time independent}).


\begin{figure}[t]
\begin{center}
\includegraphics[scale = 1]{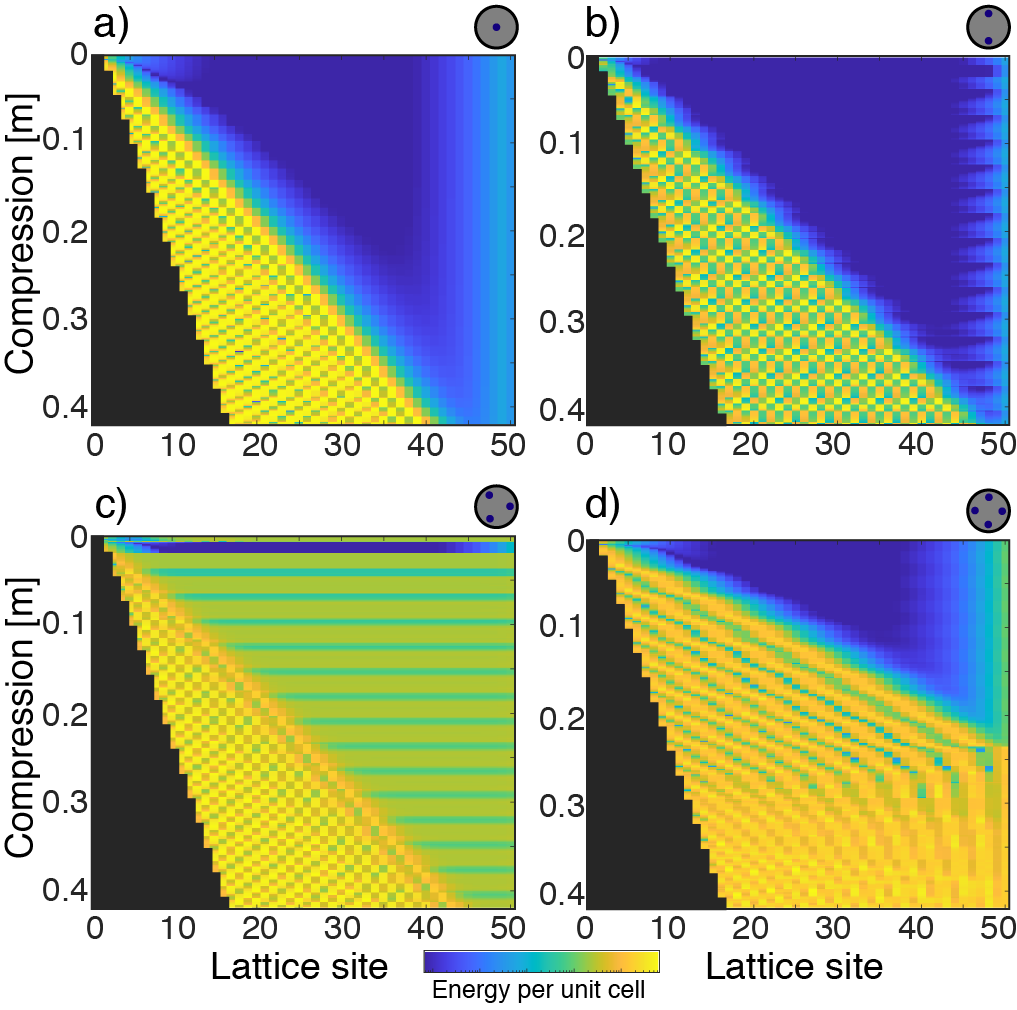}
\caption{\label{fig:transition_num}
\textbf{Time-independent transition wave.} Numerically obtained energy per unit cell as the lattice is compressed for a) one (b) two (c) three and (d) four embedded magnets per disk. }
\end{center}
\end{figure}

\begin{figure*}
\begin{center}
\includegraphics{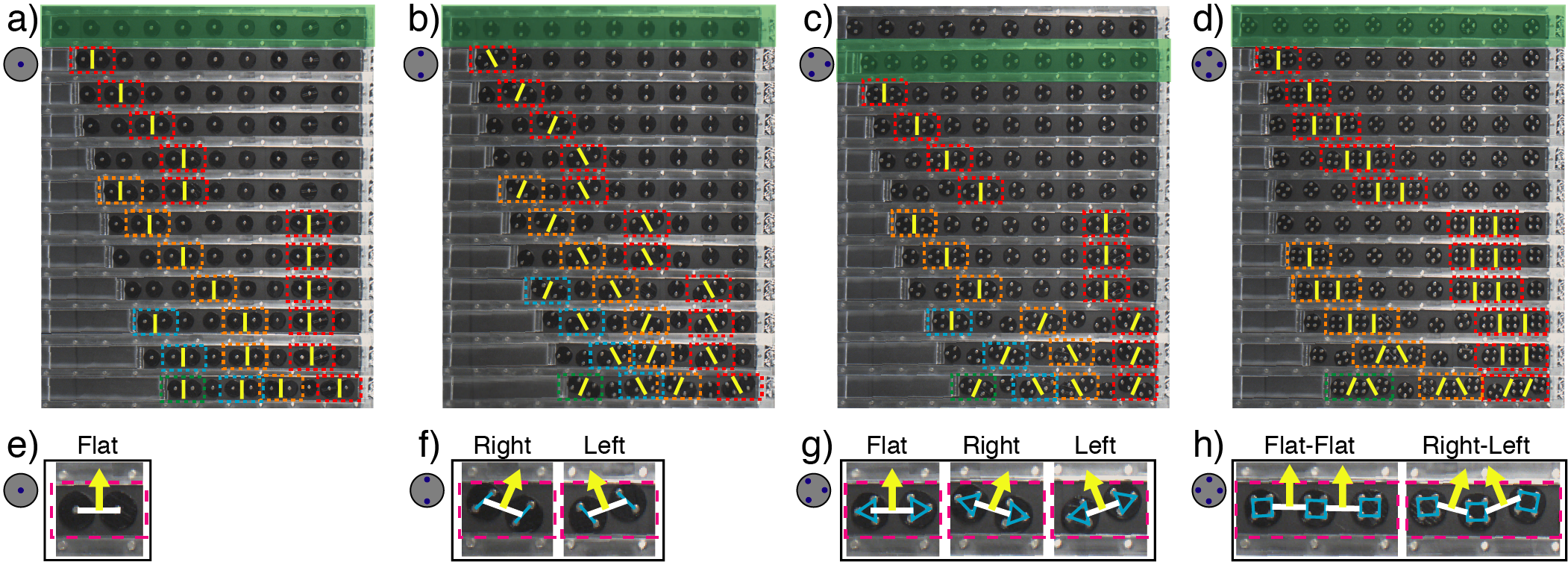}
\caption{\label{fig:transition_exp} \textbf{(a-d)Defect nucleation and propagation.} Step-wise deformation of ten levitated disks with 1-4 embedded magnets per disk. The defects are marked with with the same color dashed box as they propagate through the material. (e-h) The different types of defects arising for various disk types. The perfectly aligned state is highlighted with a green solid rectangle (top picture in each panel except for the frustrated state in c). The normals are highlighted in yellow throughout all panels.}
\end{center}
\end{figure*}

We numerically simulate the compression of a 50 disk chain for each disk type using the Verlet method (Fig.\ref{fig:transition_num}). The disks start at random positions (and phase angle in cases of more than one magnet per disk) within their potential wells. After reaching equilibrium, we periodically increase the load applied at the leftmost disk with intermittent pauses allowing the disks to reassemble. In the case of one magnet per disk,  the material starts at an equilibrium state with all disks experiencing the same repulsion forces, except those at the boundary (Fig.\ref{fig:transition_num}a). As the load increases, the leftmost disk escapes its potential well, snaps into the next well, and consequently creates the first defect in the chain. This takes place once the disk overcomes the energy barrier formed by the vertical pair of boundary magnets. When the load increases further, the second disk snaps into the next potential well and the defect propagates further down the chain as another defect nucleates at the load interface. The transition wave propagating into the chain creates a checker-like energy pattern (Fig.\ref{fig:transition_num}a). The same phenomenon arises when compressing a chain of disks with two and four embedded magnets (Fig.\ref{fig:transition_num}b,d). In the case of two-magnets per disk, the energy pattern is distinctly different from the previous pattern (for 1 magnet per disk) and the transition wave propagates further down the chain at the same rate of compression as panel a. However, the transition wave does not reach the end of the chain in either case. At four-magnets per disk, the transition wave reaches the end of the chain at ($\approx$ 0.25m) compression, significantly changing the emerging energy pattern. In the case of 3 magnets per disk, the equilibrium positions before deformation have a high energy state due to the frustration-induced disorder. This state is elucidated by the thin green line at top of Figure \ref{fig:transition_num}c, which is consistent with our previous experimental observation (Fig.\ref{fig:Dispersion}k). Surprisingly, the frustration-induced disorder vanishes after applying a small load to the chain. This compression ($\approx$ 0.01m) causes all the disks to align perfectly with a minimum energy state (a thin horizontal dark blue line in Figure \ref{fig:transition_num}c). The self-alignment, or this emergent \textit{deformation-induced order} happens every time the chain is compressed by a full unit cell length (i.e., 0.028m), appearing as flat blue lines in Figure \ref{fig:transition_num}c.

To experimentally verify our numerical observations, we consider a chain of 10 free-floating disks within the same magnetic boundary (Fig.\ref{fig:Dispersion}). We apply a load at the leftmost disk, at equilibrium (Fig.\ref{fig:transition_exp}a-d), causing the disk to snap into the next potential well. This nucleates the first defect (highlighted with dashed red boxes in Figure\ref{fig:transition_exp}a-d). In the case of 3 magnets per disk, we experimentally observe the emergent \textit{deformation-induced order}. All the ordered lattices are highlighted green in Figure \ref{fig:transition_exp}a-d. As the load increases, the defect migrates further down the chain and a new defect arises (highlighted with dashed orange boxes in Figure \ref{fig:transition_exp}a-d). As more defects nucleate, a pattern of alternating defect and defect-free cells emerges. This is congruent with the energy patterns observed numerically (Fig.\ref{fig:transition_num}).

We observe different defect topologies depending on the disk type. For disks with one embedded magnet, we identify one defect type as: two disks, aligned horizontally, confined within the same potential well (Fig.\ref{fig:transition_exp}e). In the case of two magnets per disk, we observe two distinct defects with two disks per potential well. The normal to the line connecting the centers of the two defect disks points either left or right (Fig.\ref{fig:transition_exp}f). In the case of three magnets per disk, we observe a combination of these three defect topologies (i.e., flat, pointing left or pointing right) (Fig.\ref{fig:transition_exp}g). In the case of four magnets per disk, the defect is composed of three disks, not two, sharing two adjacent potential wells. There is either a horizontal alignment of the three disks or a mix of left and right pointing normals (Fig.\ref{fig:transition_exp}h). 

\begin{figure}
\begin{center}
\includegraphics{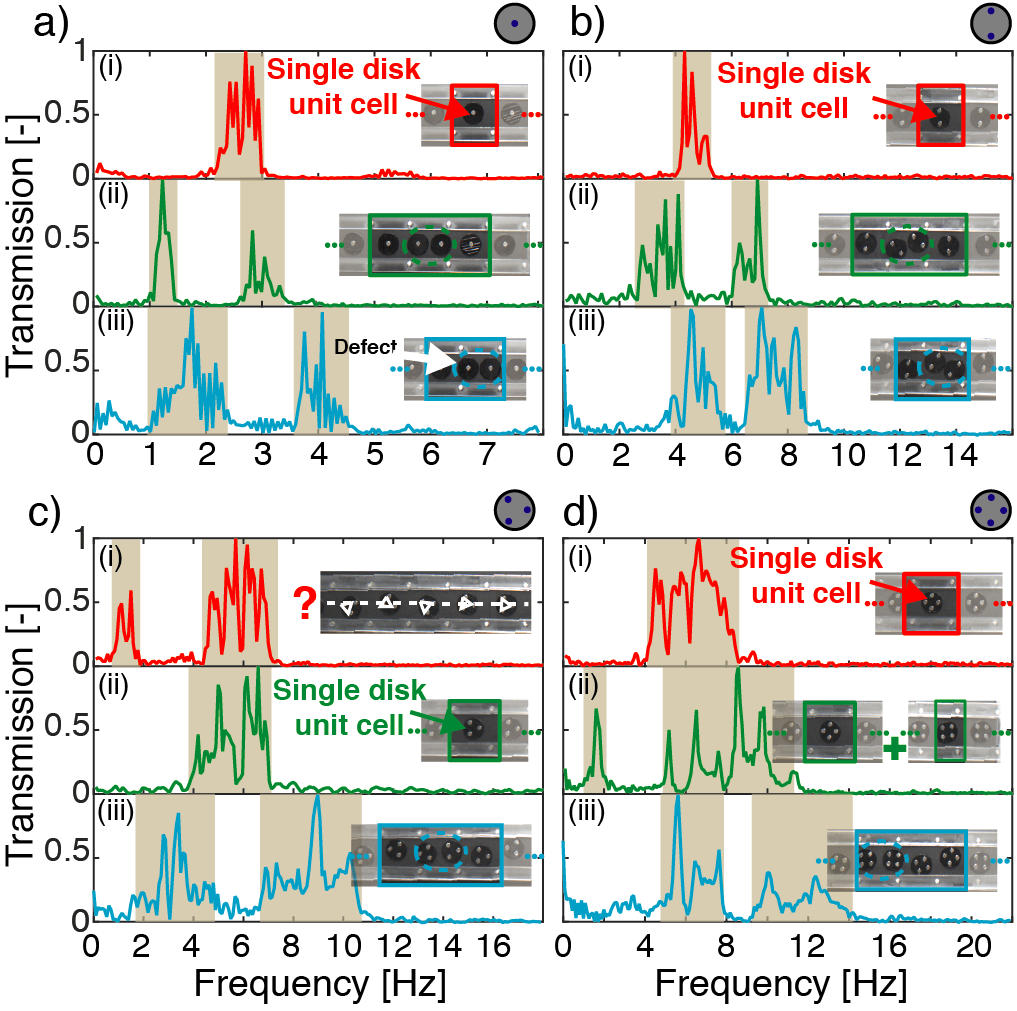}  
\caption{\label{fig:tunable} \textbf{Tunable wave propagation.} (a-d) Experimentally measured transmission in a lattice with 1-4 embedded magnets per disk. (i) Transmission through the uncompressed configuration for each disk type. (ii-iii) Transmission within the lattice after deformation. The insets show the corresponding basic building block in each case.}
\end{center}
\end{figure}

The deformation-induced pattern transformation displays a distinct topological signature with different disk couplings. This translates to a change in material properties as patterns evolve. We experimentally test the dynamical characteristics of the emerging patterns by exciting three of the newly formed lattices for each of the disk types (Fig.\ref{fig:tunable}). For reference, we first harmonically excite the undeformed self-aligned chain of ten magnets with a chirp signal between 0.2-35 Hz and record the transmission through the lattice (1D-FFT). The most left disk is then compressed until a new pattern emerges across the ten disks. The chain is then excited with the same chirp signal. The experiment is repeated for two emerging patterns (panels (ii,iii) in Fig.\ref{fig:tunable}) in addition to the pattern of the undeformed disks (panels (i) in Fig.\ref{fig:tunable}). The basic repeating pattern in each case is also included (Fig.\ref{fig:tunable}-insets).

With the one-magnet per disk arrangement, the reference pattern has a single longitudinal transmission band  (Fig.\ref{fig:tunable}a(i)). This single band turns into two distinct bands when the pattern is a defect cell sandwiched between two defect-free cells (Fig.\ref{fig:tunable}a(ii)).  As the load increases and the pattern turns into a single defect-free cell and a defect cell, both bands widen and shift to higher frequencies (Fig.\ref{fig:tunable}a(iii)). The same phenomenon takes place for disks with two embedded magnets. The reference assembly with perfectly aligned disks transitions into a defect sandwiched between two defect-free cells. The pattern then transforms into a single defect-free and a defect cell, with the emergence of two wider transmission bands at higher frequencies (Fig.\ref{fig:tunable}b(i-iii)). In the three magnets per disk case, the initial pattern is frustrated with no long range order. As observed earlier (Fig.\ref{fig:Dispersion}g), two transmission bands are present (Fig.\ref{fig:tunable}c(i)). Once the lattice is compressed, the deformation-induced order emerges and the low frequency disorder-induced band disappears (Fig.\ref{fig:tunable}c(ii)). As the load increases, the pattern transforms further into a defect cell sandwiched between two defect-free cells (Fig.\ref{fig:tunable}c(iii)). Finally, At four magnets per disk, the lattice transforms from aligned disks in a 0$^{\circ}$ phase angle into two neighbouring lattices with 0$^{\circ}$ and 45$^{\circ}$ phase angles (Fig.\ref{fig:tunable}d(ii)). The difference in transformation for this disk type is a combination of both the number of disks within the defect (i.e., 3 instead of 2) and the small number of disks considered in the experiment (i.e., only 10 disks). As the load increases, two transmission bands emerge as the pattern transforms into an alternation of two defect-free cells and one defect cell (Fig.\ref{fig:tunable}d(iii)).


In summary, we consider free-floating magnetic disks within a magnetic boundary with periodic potentials. The levitated disks perfectly align in the case of 1, 2 and 4 magnets per disk. In the case of 3 magnets per disk, the global lattice assembly shows no long range order due to the presence of geometric frustration. Despite the clear disorder, the transmission within the lattice closely resembles that of a perfectly ordered lattice. For all different disk types, we further transform the self-alignment of the disks by applying a load at one end of each emerging lattice. We observe the nucleation and propagation of different defect types (depending on the number of magnets per disk) within the lattices, giving rise to  \textit{time-independent} stable transition waves due to instabilities. Remarkably, in the presence of frustration (i.e., 3 magnets per disk), the applied load briefly introduces a stable order to the assembly. By further deforming the lattices, new patterns emerge across all disk types. We harness these emerging patterns to tune the wave propagation characteristics of the soft material. Our findings shed light on the understanding of deformation-induced pattern transformations, particularly the rise of frustration-induced disorder and deformation-induced order. The nature of the localized and stable defect propagation, in addition to the nonlinear coupling potentials, can be harnessed in designing the next generation of soft materials for applications ranging from sound manipulation to soft robotics.


\begin{acknowledgments}
This work is supported by the University of Connecticut's start-up fund (ORB).
\end{acknowledgments}


%

 \newpage


 \newpage
 \beginsupplement

\begin{widetext}
\hspace{-3mm}\Large{\textbf{Supporting Information: \\}}
\Large{\textbf{Exploiting Localized Transition Waves to Tune Sound Propagation in Soft Materials}\\}

\begin{figure}[b]
\begin{center}
\includegraphics{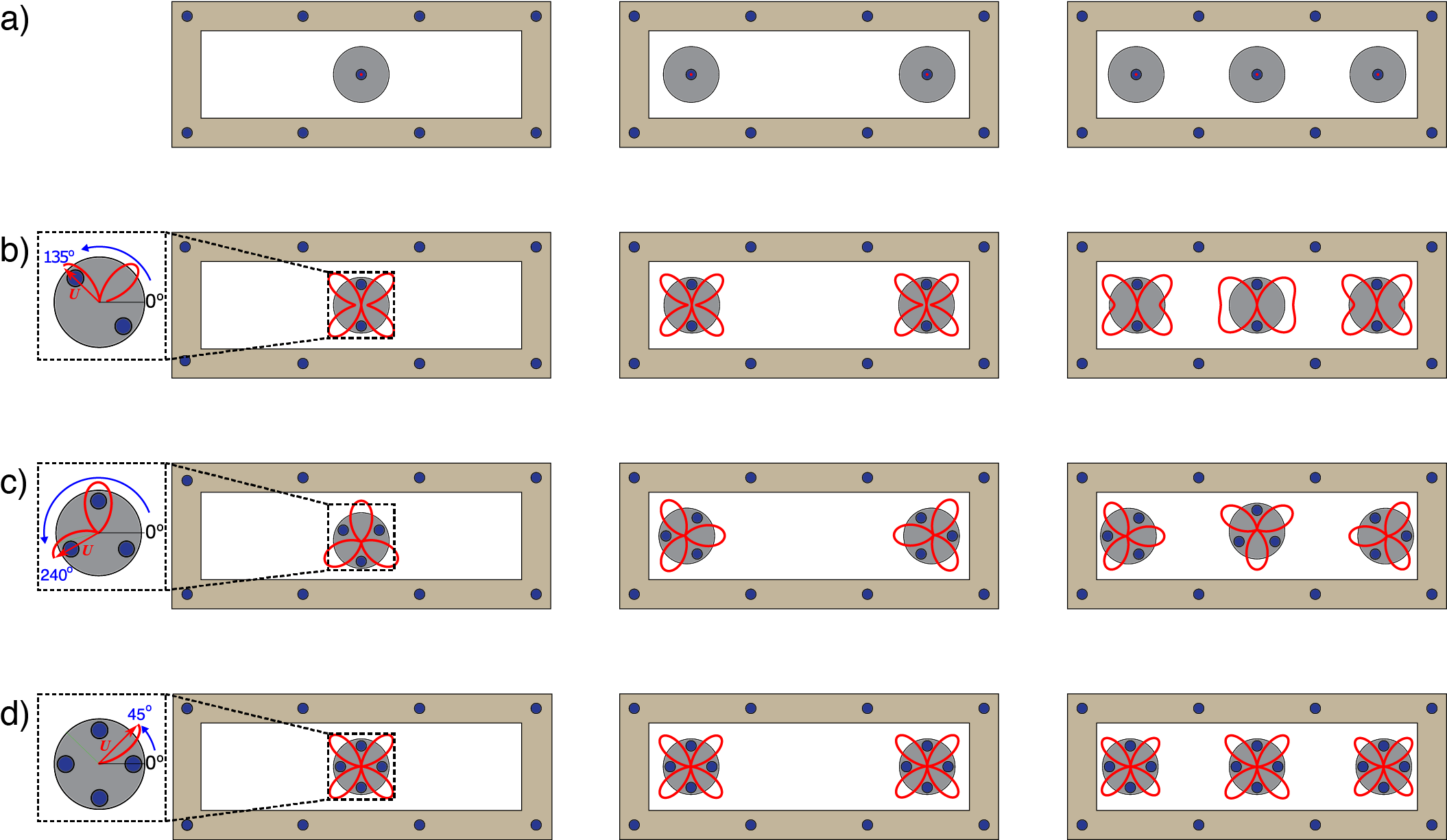}
\caption{\label{figS:orientation of disks} \large{\textbf{Minimum energy states.} Equilibrium position and orientation of disks within a fixed magnetic boundary with (a) one, (b) two, (c) three, and (d) four embedded magnets per disk. Polar plot of the energy per disk as a function of the disk phase angle superimposed on the corresponding disks.}}
\end{center}
\end{figure}

\textbf{Minimum energy states:}
The equilibrium positions of one, two, and three disks within a magnetic boundary are determined both numerically and experimentally. Each disk is first positioned at a random location within its own potential well inside the boundary. The disks move in both $x$ and $y$ directions and rotate freely to find their lowest energy state. The numerically calculated minimum energy state for each case is presented in Figure \ref{figS:orientation of disks}. After the disk reaches its equilibrium position, the disk is rotated 360$^\circ$ and the change of energy as a function of orientation angle is overlaid on top of each disk as a polar plot: 
\begin{equation}
U = -\int_{0}^{\theta_{d}}r\times \sum_{n=1}^{N} \left\{\Sigma F_{y}\cdot cos(\theta_{d}+\frac{2\pi n}{N})-\Sigma F_{x}\cdot sin(\theta_{d}+\frac{2\pi n}{N})\right \}d\theta_{d},
\end{equation}

where $\theta_{d}$ is the phase angle of the disk and $\Sigma F_{x}$ and $\Sigma F_{y}$ are the sum of forces in the $x$ and $y$ directions, respectively, on magnet $n$.

\begin{figure}[b]
	\begin{center}
		\includegraphics[width= \textwidth]{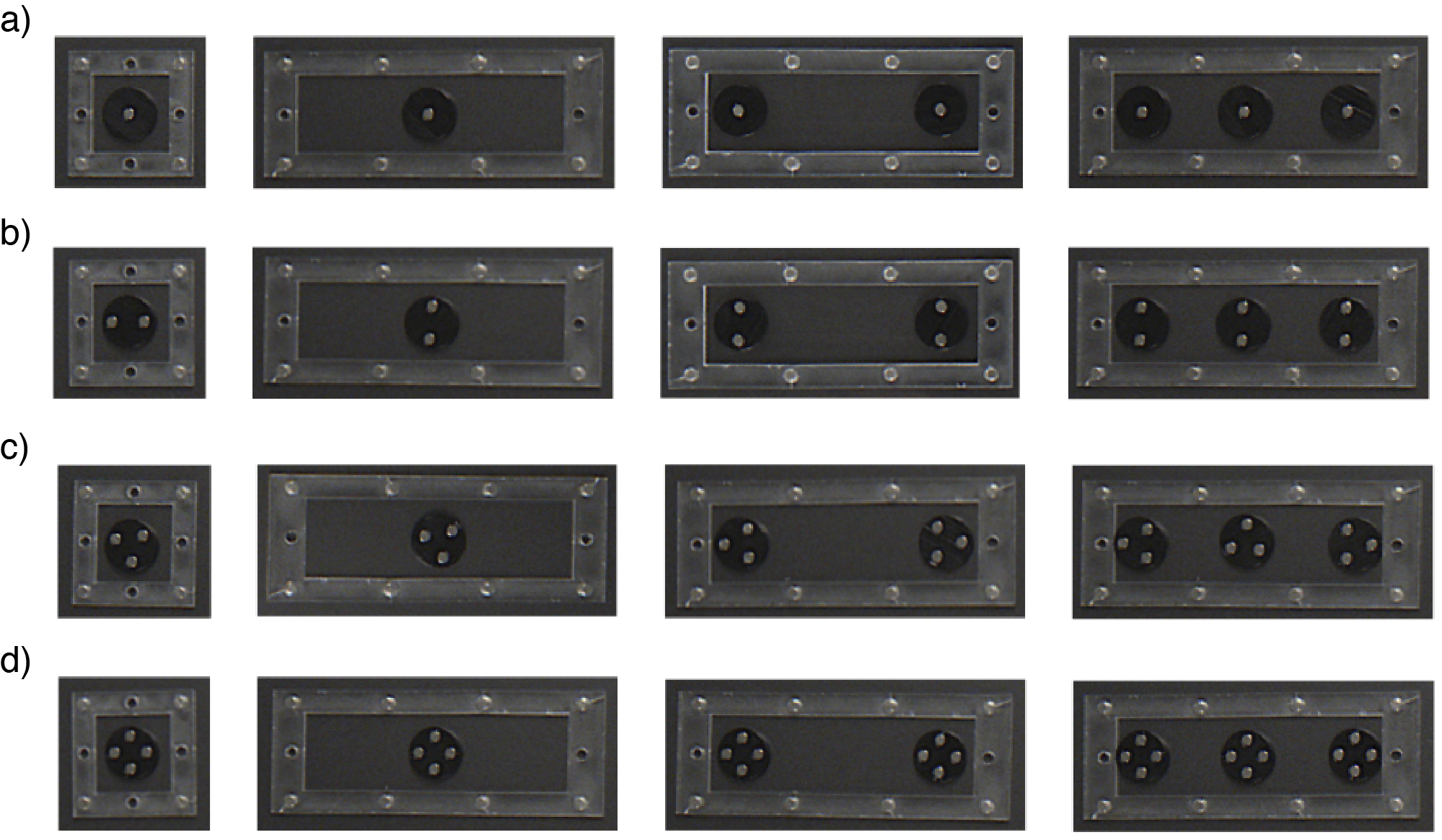}
		\caption{\label{fig:supp. experimental orientations} \large{\textbf{Experimental disk minimum energy states.} Equilibrium position and orientation of disks within a fixed magnetic boundary with (a) one, (b) two, (c) three, and (d) four embedded magnets per disk.}}
	\end{center}
\end{figure} 

In the case of one magnet per disk, the minimum energy state is at the center of the potential well for one, two, or three disks. Given the symmetry of the disk, the energy minima is independent from the disk's phase angle (Fig. \ref{figS:orientation of disks}a). In the case of two magnets per disk, the minimum energy state is located at the middle of the potential well with a 90$^\circ$ (vertical) phase angle. The same vertical orientation represents the equilibrium positions for one, two, and three disks, separately,  within the magnetic boundary (Fig. \ref{figS:orientation of disks}b). It is important to note the change in the energy polar plot when \textit{all} three disks are present within the boundary (Fig. \ref{figS:orientation of disks}b-right). In the case of a single disk with three embedded magnets, the minimum energy state is not located at the middle of the potential well, but is rather slightly higher or lower than the center point along the y-axis. When two disks with three embedded magnets are positioned within the boundary, they retain an opposite phase angle in order to minimize their own energy. By adding a disk at the central potential well, (i.e., when three disks are enclosed by the boundary), there exists no unique phase angle or location for all disks to minimize their own energy as well as the total energy of the assembly, hence the system is geometrically frustrated (Fig. \ref{figS:orientation of disks}c). In the case of four magnets per disk, the minimum energy state is located at the middle of the potential well with a 0$^\circ$ (plus sign) phase angle. The same plus sign orientation represents the equilibrium positions for one, two, and three disks, separately,  within the magnetic boundary (Fig. \ref{figS:orientation of disks}d). The energy polar plots of all disks do not change when the third disk is added within the boundary (Fig. \ref{figS:orientation of disks}d-right), in contrast to the case of two magnets per disk.

\textbf{Experimental methods:}
The fixed boundaries are laser cut out of a transparent acrylic plastic sheet using a Full-spectrum 24 pro-series laser cutter. The spacing between the embedded magnets (from magnet center to center) in the boundary is 28 mm in both the $x$ and $y$ directions. The width of the channel in the $y$ direction is 20 mm. The disks are fabricated out of VeroBlack material using an Object 3D printer (Connex 500). The fabricated disks have a radius $r_{disk} = 7.5$ mm. Both the boundary and the disk magnets have a radius of $r_{magnet} = 1.5$ mm.  The disks are placed within the boundary so that the magnets are oriented with their north pole facing upwards. The fixed boundary magnets have the same magnetic  orientation. To minimize friction, we attach a glass cover \textcolor{black}{slide} at the bottom of each disk and float them on an air bearing (New way S1030002).

To experimentally validate the potential energy calculations in  Figure \ref{figS:orientation of disks}, we enclose one, two, and three disks of each disk type (i.e., with 1-4 magnets) within a fixed boundary. The disks start in random positions relative to the boundary and to each other. Then, we activate a layer of laminar air flow beneath the surface of the disks to allow them to float freely (i.e., levitating the disks similar to the arcade game ``Air Hockey"). As predicted numerically, the disks with 1, 2, and 4 embedded magnets self-align in a perfectly ordered lattice (Fig. \ref{fig:supp. experimental orientations}).

To experimentally determine the dynamical properties of our lattices, we excite the rightmost end of the lattice with a mechanical shaker (Br\"{u}el and Kj\ae r 4180) and a function generator (Keysight Technologies 33512B). The excitation takes place as a chirp signal, ranging from 0.2 Hertz  - 35 Hertz. The motion of the disks is captured using a computer vision camera (Blackfly S USB3) and the resulting images are analyzed using the digital image correlation software (DICe). A fast Fourier transform (FFT) is applied to the displacement signals and the transmitted frequency ranges are observed. The FFT graphs depict the ranges of propagation and attenuation of mechanical waves through the lattice. \textcolor{black}{We obtain our experimental dispersion curves by exciting the lattices using a broad frequency spectrum. Then track the motion of discrete points along the wave propagation direction (in our case the lattice -made out of disks- is already discrete). The resulting matrix (which includes the displacement time-history of each particle) can be transformed to the presented dispersion curves by applying the fft2 command in Matlab.}

\begin{figure}
\begin{center}
\includegraphics[width = \textwidth]{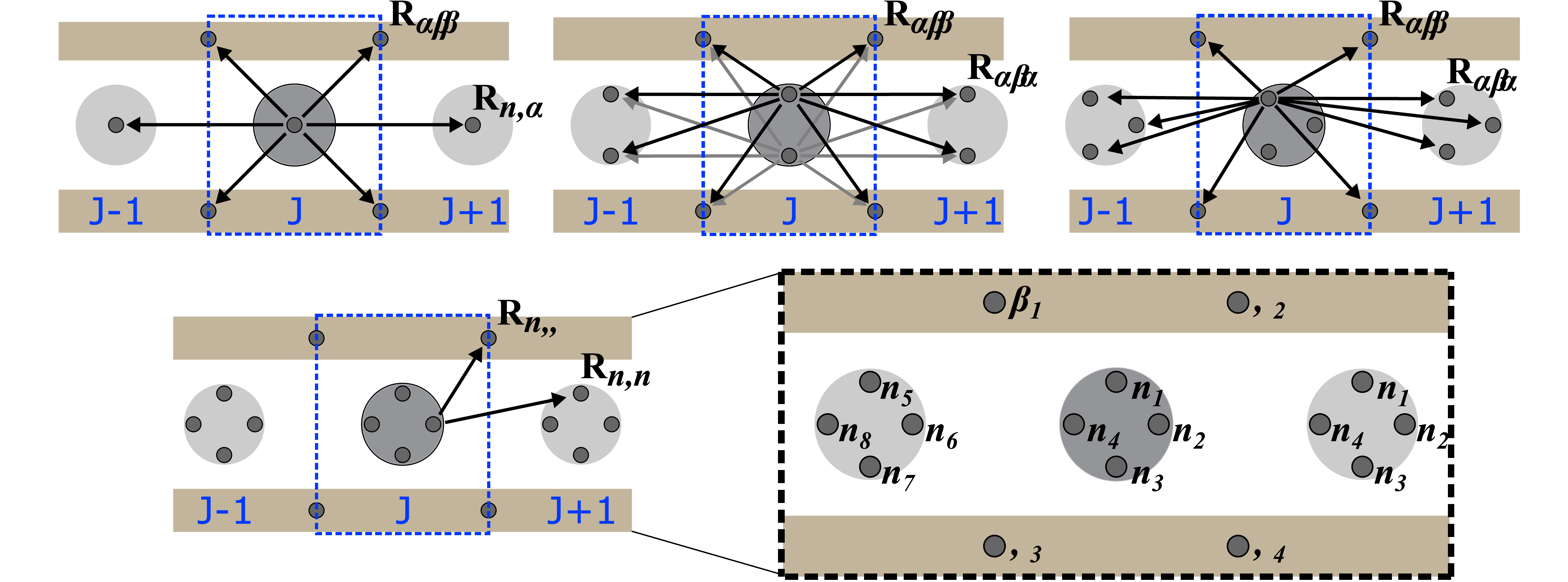}
\caption{\label{fig:force vectors} \large{\textbf{Disk Unit Force Vectors.} Location and coupling of the magnetic forces used in determining the stiffness matrix of a lattice with disks containing 1, 2, 3, or 4 embedded magnets.}}
\label{suppfig:vectors}
\end{center}
\end{figure}

\textbf{Dispersion Analysis:}
The dispersion relation of each system is calculated by solving the eigenvalue problem \begin{equation}
    [-\omega^2\textbf{M}+\mathbf{K}(\boldsymbol\kappa)] \boldsymbol{\phi} = 0
    \label{eqn:eigen}
\end{equation} 
where $\omega$ is the frequency, $\kappa$ is the wavenumber, $\boldsymbol\phi = [u~v]^{T}$ is the Bloch displacement vector in the $x$ and $y$ directions $\boldsymbol {M}$ is the mass matrix:  
\begin{equation}
\boldsymbol {\mathbf{M}} = \begin{bmatrix}
m_i & 0\\
0 & m_i
\end{bmatrix}
\end{equation}
where $m_i$ is the mass of the disk with $i$ embedded magnets ($i= 1,...,4$). The stiffness matrix is a function of wave number as:

\begin{multline}
\mathbf{\textbf{K}\left (\kappa   \right )}=\sum_{n=1}^{N} \Biggl( 
\sum_{\alpha=1}^{2N} \biggl[ f_{,d}(d_{n,\alpha})\textbf{e}_{n,\alpha}\otimes \textbf{e}_{n,\alpha}\left [ cos(\kappa \cdot \textbf{R}_{n,\alpha})-1 \right ]\\
+ \frac{f(d_{n,\alpha})}{(d_{n,\alpha})}(\textbf{I}-\textbf{e}_{n,\alpha}\otimes \textbf{e}_{n,\alpha})\left [ cos(\kappa\cdot \textbf{R}_{n,\alpha})-1 \right ] \biggr]\\ 
- \sum_{\beta=1}^{4}\biggl[ f_{,d}(d_{n,\beta})\textbf{e}_{n,\beta}\otimes \textbf{e}_{n,\beta} + \frac{f(d_{n,\beta})}{(d_{n,\beta})}(\textbf{I}-\textbf{e}_{n,\beta}\otimes \textbf{e}_{n,\beta})\biggr] \Biggr)
\end{multline}

where $\alpha$ is the index of the magnets in neighboring unit cells (e.g. $j+1$, $j-1$), $\beta$ is index of the boundary magnets, $N$ is the number of magnets in a single disk, $n$ is the index of the magnet in unit cell $j$, $\boldsymbol{R_{n,\alpha}}$ is the unit vector between magnet $n$ and $\alpha$ and $\otimes$ is the dyadic product. Figure \ref{suppfig:vectors} shows a visualization of the unit vectors between unit cells for the different disk types. \textcolor{black}{The repulsive forces between the magnets are modeled as an inverse power law $f(d) = A d^{\gamma}$} with $f_{,d}(d)$ representing the function's first derivative \textcolor{black}{and $\gamma = -4$. In the instance of dipole-dipole interaction, $A = 3\mu \beta ^2/4\pi= 9.617 \times 10^{-11}$ where $\mu$ is the permeability of air and $\beta$ is the magnetic moment \cite{watkins2020demultiplexing}.}\\ 

\textbf{Numerical Analysis.}
The compression of the lattice and the transformation of the patterns within the lattice are also considered numerically. To simulate the nucleation and propagation of defects in the different lattices, four cases of 10 disks confined to a boundary potential are considered. The equation of motion for each disk is integrated in time using the Verlet method \cite{press2007numerical}. A small amount of damping is introduced to each of the disks while an increment of 1 mm of compression over a 5 second period occurs over 10 seconds intervals; the additional 5 seconds are included to allow for the disks to stabilize. Depending on both the number and arrangement of magnets within each disk, the configuration of disks in the lattice can restructure themselves into a new ordered pattern based on the ratio of elements to potential wells (Fig.\ref{fig:compression steps}). 

\begin{figure}[t]
\begin{center}
\includegraphics[width= 0.97\textwidth]{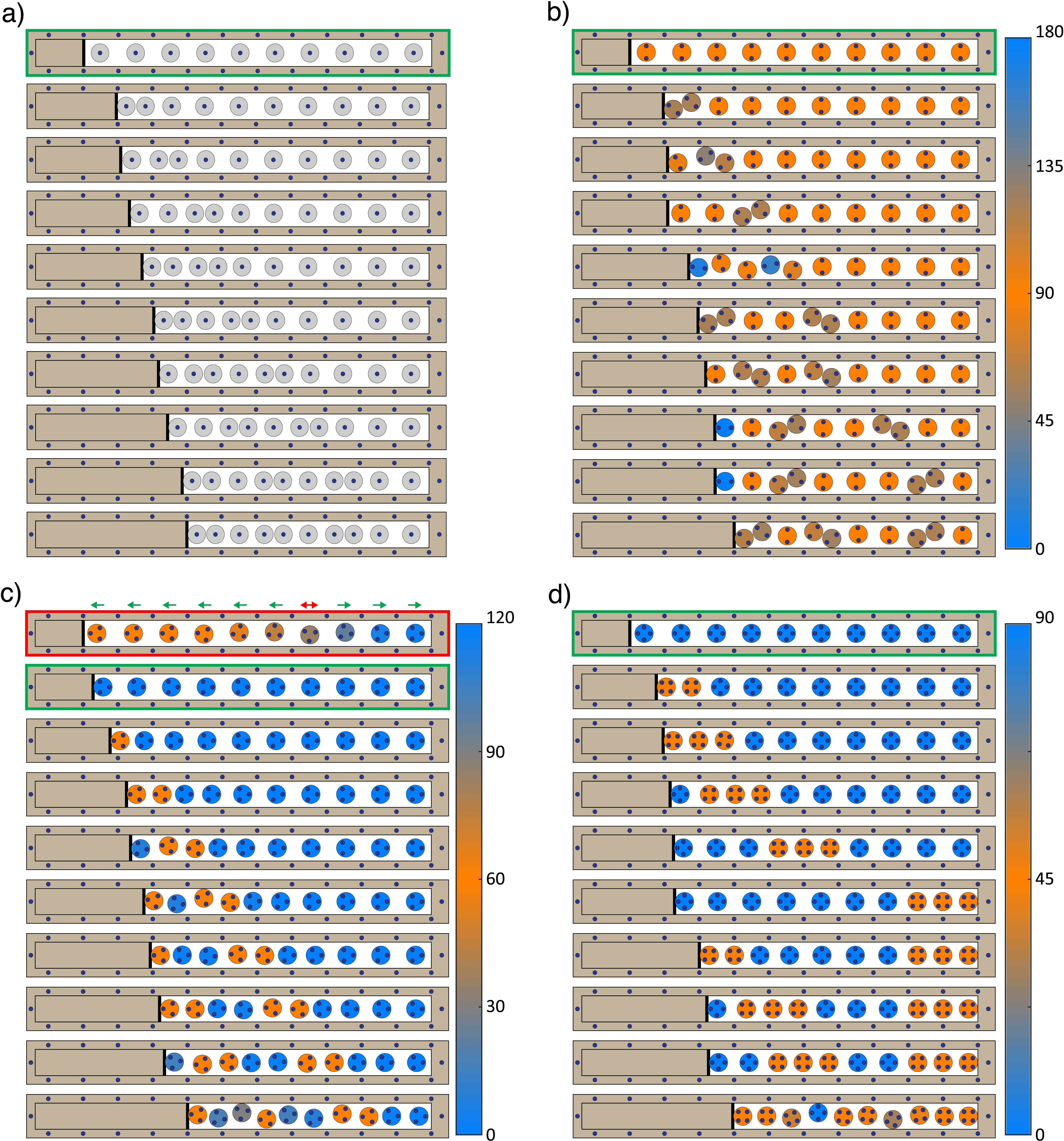}
\caption{\label{fig:compression steps} \large{\textbf{Numerical Defect Nucleation and propagation.} Compression of 10 disks confined to a fixed magnetic boundary with (a) one, (b) two, (c) three, and (d) four embedded magnets per disk.}}
\end{center}
\end{figure}

In the case of one magnet per disk, the uncompressed system is initially ordered. As the compression begins, defects of two disks per potential well form at the left end and travel right. This behavior continues and eventually fills the entire system with an alternating pattern of one and two disks per potential well (Fig \ref{fig:compression steps} a). For disks with multiple magnets, the color of the disk corresponds to its phase angle (Fig. \ref{fig:compression steps} b-d). In the case of two magnets per disk, all disks align vertically (90$^\circ$ phase angle) shortly after the simulation starts. As the disks are compressed from the right, a defect of two disks occupying the same potential well nucleates.  Similar to our observation in experiments, the defect propagates down the lattice and new ones emerge at the compression front. The defect types that emerge are similar to the ones identified experimentally in the main text of the paper (i.e., right or left leaning normals to the line connecting the disk centers). The defect cells can be easily identified with the brown phase color of the two disks within the same well. In the case of three magnets per disk, the uncompressed system is already disordered (note the change in color between both ends of the lattice) and only becomes ordered after a slight compression  (Fig. \ref{fig:compression steps}c). Although the uncompressed system is not ordered in the same way as the systems with one, two, and four magnets, the creation and propagation of defects still occur in a similar manner. The defect types within the numerical simulations are also similar to those observed in the experiments. Alternating orange and blue disks within the same well show a flat defect and gray off-center disks show the right or left defects (Fig. \ref{fig:compression steps}c). In the case of four magnets per disk, an ordered self-assembly is a natural, unforced occurrence, similar to the assembly of disks with 1 and 2 embedded magnets. As the system is compressed, defects form and travel down the length of the lattice. Unlike the instances of 1, 2, and 3 magnets per disk, the defects that travel through the lattice in the case of 4 embedded magnets consist of three disks per defect. The defect travels through the lattice in a similar manner, with more defects emerging as the level of compression is increased (Fig. \ref{fig:compression steps}d).

\begin{figure}
	\begin{center}
		\includegraphics[width= \textwidth]{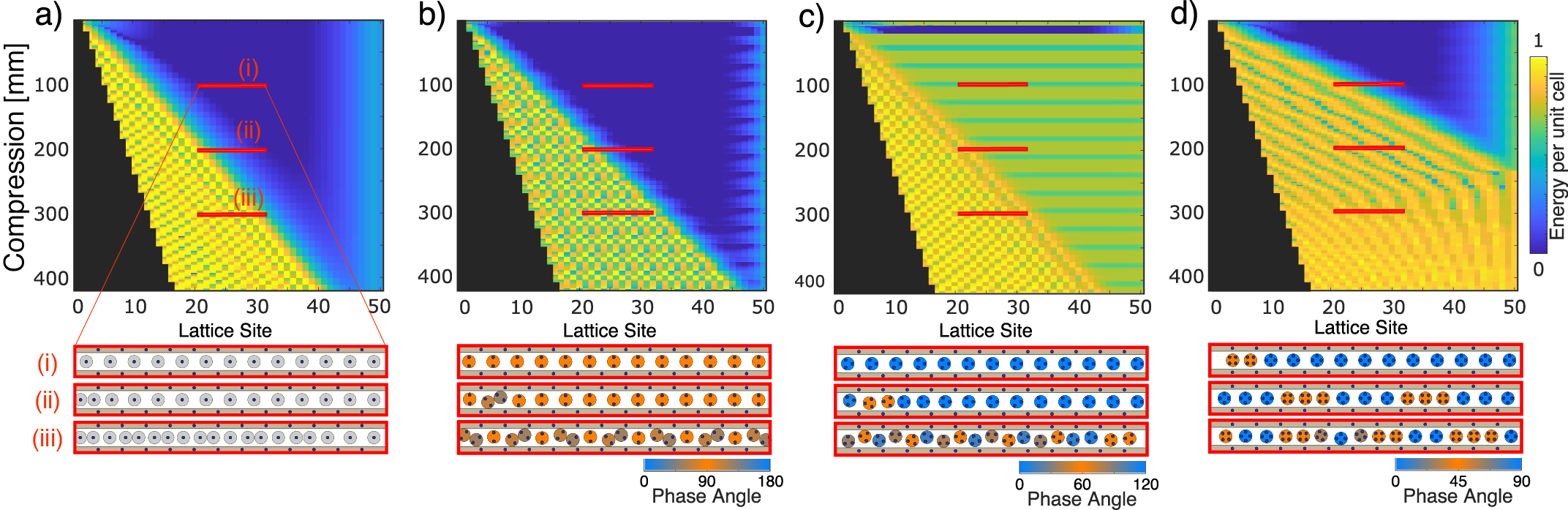}
		\caption{\label{figS:transition_num}
			\large{\textbf{Time-independent transition wave.} Numerically obtained energy per unit cell as the lattice is compressed for (a) one (b) two (c) three and (d) four embedded magnets per disk.}}
	\end{center}
\end{figure}

To test the generality of the nucleation and propagation of defects apart from boundary (edge) effects, we simulate the deformation of longer lattices confined to the boundary potentials. We consider a lattice made out of 50 disks for each disk type (i.e., 1-4 embedded magnets). To visualize defects in a much larger lattice, we use the energy per disk within a unit cell to locate defects within the lattice. Unit cells with a relatively higher energy typically contain more than 1 disk (i.e., a defect exists). We represent the energy per unit cell (i.e., the four boundary magnets) as:

\begin{equation}
E_{UC} = \sum_{\beta = 1}^{4}  \sum_{k = 1}^{D}\sum_{j = 1}^{N} | {A d_{\beta k_{j}}^{\gamma}}|
\end{equation}

where $D$ is the number of free-floating disks within the unit cell boundaries, $N \in [{1,2 \dots 4}]$ is the number of magnets per disk, $d_{\beta k_j}$ is the distance between the position of magnet $j$ in disk $k$ and magnet $\beta$ in the fixed boundary, $A = 3\mu \beta ^2/4\pi= 9.617 \times 10^{-11}$, where $\mu$ is the permeability of air and $\beta$ is the magnetic moment. The compression of the lattices is considered for the length of 15 unit cells (i.e., 420 mm). The energy level changes are plotted in Figure \ref{fig:transition_num}. Supplementary Figure \ref{figS:transition_num} gives a closer look at what the energy evolution mean physically. Each red horizontal line corresponds to the respective set of thirteen unit cells within the lattice (\ref{figS:transition_num}(i-iii)). In each case, as the defect begins to travel down the lattice, the calculated energy corresponds to the number of disks within a single potential well and the disk's phase angles. The blue coloring of the energy graphs corresponds to the ordered alignment of the disks. As the disks begin to rotate, changing their phase angles, the energy differences are apparent as the blue turns to a green, orange, or yellow color. Supplementary Figure \ref{figS:transition_num}(c) shows the rightmost end of the lattice containing 3 embedded magnets going through an alternating energy cycle as the disks rotate in and out of their perfectly ordered state.

 \newpage
\end{widetext}

\end{document}